\documentclass[manuscript,authorversion,nonacm,screen]{acmart}
\AtBeginDocument{%
  \providecommand\BibTeX{{%
    \normalfont B\kern-0.5em{\scshape i\kern-0.25em b}\kern-0.8em\TeX}}}

\usepackage{lipsum}
\usepackage{soul}
\usepackage{xargs}
\usepackage{natbib}
\usepackage{amsmath}
\usepackage{stackengine}
\usepackage{setspace}
\usepackage{graphicx, tabularx}

\usepackage[nolist]{acronym}
\begin{acronym}
    \acro{AI}{Artificial Intelligence}

    \acro{HCI}{Human-Computer Interaction}

    \acro{IRR}{Inter-Rater Reliability}

    \acro{LLM}{Large Language Model}

    \acro{NLP}{Natural Language Processing}

    \acro{QDA}{Qualitative Data Analysis}

\end{acronym}

\setstackgap{S}{1pt}
\newcommand{\hltop}[3]{\sethlcolor{#3}\stackengine{\stackgap}{\hl{#1}}{\color{#3}\texttt{\footnotesize #2}}{O}{l}{\quietstack}{\useanchorwidth}{S}}

\definecolor{lightcoral}{rgb}{0.94, 0.5, 0.5}
\definecolor{lightcornflowerblue}{rgb}{0.6, 0.81, 0.93}
\definecolor{lightfuchsiapink}{rgb}{0.98, 0.52, 0.9}
\definecolor{lightgreen}{rgb}{0.55, 0.72, 0.16}
\definecolor{lightsalmon}{rgb}{1.0, 0.63, 0.48}
\definecolor{lightgray}{rgb}{0.73, 0.73, 0.73}
\definecolor{lightpastelpurple}{rgb}{0.69, 0.61, 0.85}
\definecolor{lightseagreen}{rgb}{0.13, 0.7, 0.67}

\newcolumntype{C}{>{$}c<{$}} 

\newcommand{\eg}{e.\,g.}
\newcommand{\etal}{et~al.\@\,}

\begin{document}

\title[Exploring Human-AI Concordance in Qualitative Coding]{Decoding Complexity: Exploring Human-AI Concordance in Qualitative Coding}


\author{Elisabeth Kirsten}
\orcid{0009-0003-0680-8916}
\email{elisabeth.kirsten@mpi-sp.org}
\affiliation{%
  \institution{Max Planck Institute for Security and Privacy}
  \city{Bochum}
  \country{Germany}
}

\author{Annalina Buckmann}
\orcid{0000-0002-7959-9743}
\affiliation{%
  \institution{Ruhr University Bochum}
  \city{Bochum}
  \country{Germany}
}

\author{Abraham Mhaidli}
\orcid{0000-0002-9519-245X}
\affiliation{%
  \institution{Max Planck Institute for Security and Privacy}
  \city{Bochum}
  \country{Germany}
}

\author{Steffen Becker}
\orcid{0000-0001-7526-5597}
\affiliation{%
  \institution{Ruhr University Bochum}
  \city{Bochum}
  \country{Germany}
}
\affiliation{%
  \institution{Max Planck Institute for Security and Privacy}
  \city{Bochum}
  \country{Germany}
}

\renewcommand{\shortauthors}{Kirsten \etal}

\begin{abstract}
Qualitative data analysis provides insight into the underlying perceptions and experiences within unstructured data.
However, the time-consuming nature of the coding process, especially for larger datasets, calls for innovative approaches, such as the integration of Large Language Models (LLMs).
This short paper presents initial findings from a study investigating the integration of LLMs for coding tasks of varying complexity in a real-world dataset. 
Our results highlight the challenges inherent in coding with extensive codebooks and contexts, both for human coders and LLMs, and suggest that the integration of LLMs into the coding process requires a task-by-task evaluation.
We examine factors influencing the complexity of coding tasks and initiate a discussion on the usefulness and limitations of incorporating LLMs in qualitative research.
\end{abstract}






\maketitle

\section{Introduction}
\acf{QDA} is an approach used in \acf{HCI} research to analyze unstructured data, such as interview data or responses to open-ended survey questions.
A fundamental step in \ac{QDA} is coding, the systematic process of identifying, annotating, and categorizing data segments based on recurring themes and patterns, and assigning according categories or ``codes''~\citep{braun2006using, flick2022introduction, bergin2018introduction, kuckartz2019qualitative, adams2008qualititative}. 
Coding data can be a time-consuming process, particularly when analyzing large volumes of data (\eg, tens of thousands of open-ended survey responses).
The emergence of Generative Artificial Intelligence, and in particular \acfp{LLM} (\eg, GPT-4~\cite{achiam2023gpt}, Gemini~\cite{team2023gemini}, or Llama 2~\cite{touvron2023llama}) could significantly speed up the coding process, yet their value and validity needs to be evaluated. 
\acp{LLM} have demonstrated remarkable performance in annotation tasks in zero-shot or few-shot learning scenarios where no or little labeled data is given~\citep{tornberg2023chatgpt, gilardi2023chatgpt, ziems2023can}.
As these models become integral to \ac{HCI} research, it is imperative to critically evaluate their potential benefits and risks, as well as methodological challenges associated with the integration of \acp{LLM}.
In this workshop submission, we present the preliminary results of a study scrutinizing how \acp{LLM} perform at applying tags for \ac{QDA} in coding tasks of different complexity. 
Specifically, we explore \ac{LLM}-assisted \ac{QDA} of a real-world dataset to understand how \acp{LLM} perform, how they compare to human coders, and how their performance is affected by the type of coding (\eg, semantic vs. latent themes~\citep{braun2006using}).
First, we develop a strategy for prompting \acp{LLM} to apply tags for \ac{QDA}.
We then implement this strategy with GPT-3.5~\citep{brown2020language} and GPT-4~\citep{achiam2023gpt} on interview data and compare the models' performance with human coding in three tasks that require different levels of interpretation.

\section{Research Method}
To evaluate the performance of \acp{LLM} in \ac{QDA}, we provide the model with segments of qualitative data from interview transcripts\footnote{The data for our experiments is drawn from a separate interview study (n=47) focused on users' perceptions of digital security and privacy. 
The interview data is in German, and we conduct all experiments using German data segments and prompts.}, a human-generated codebook, and the instruction to assign zero, one, or more tags to each data segment using codes from the codebook. 
We then calculate the \acf{IRR} to evaluate the agreement of the resulting coded datasets between \acp{LLM} and human coders.
We analyze three distinct themes in the data, each of which presents unique challenges and considerations for coding.

\paragraph{Coding Tasks}
The models and two human coders annotate the dataset using the same human-generated codebook per task\footnote{To facilitate the coding process and adhere to the models' token length limits, two researchers extracted relevant segments from the interview transcripts for all three coding tasks.}.
We hypothesize that the difficulty of assigning tags to interview segments is influenced by several factors, such as the length of the segments, the codebook and its length, and the background knowledge and judgment required to assign appropriate categories. 
In this context, \citet{braun2006using} classify possible themes as semantic/latent or a combination of both.
Semantic themes can be identified from the surface meanings of the data by simply looking at what a participant has said. 
Latent themes go beyond the surface meaning of the data and require interpretation of the underlying ideas and assumptions.
To analyze how the similarity of results between human coders and LLMs is affected by coding for semantic versus latent themes, we perform coding for three different coding tasks:
\begin{enumerate}
    \item \textit{Task A (Internet-connected devices):} This coding task involves identifying the Internet-connected devices that participants use. We expect that the assignment of semantic codes for Internet-connected devices is straightforward for both humans and \acp{LLM}, and requires only the identification of the entities as communicated by the interviewees. (average segment length: 118 words; codebook length: 18 codes)
    \item \textit{Task B (Apps, programs, services, and use cases):} This coding task focuses on apps, programs, and services participants use on their Internet-connected devices and for what purpose. This introduces a layer of complexity, as participants may articulate their interactions in different ways, \eg, by enumerating individual apps, grouping applications, or explicitly describing their use cases. This variability introduces a hierarchy in the data that demands both semantic and latent coding. (average segment length: 274 words; codebook length: 24 codes)
    \item \textit{Task C (Trusted sources):} This coding task explores participants' practices and sources when seeking guidance on digital security and privacy. This task goes beyond capturing the semantic content of the data, as it requires examining underlying ideas and assumptions. (average segment length: 469 words; codebook length: 32 codes)
\end{enumerate}

\paragraph{Experimental Design}
For our experiments, we use OpenAI's GPT-3.5~\citep{brown2020language} and GPT-4~\citep{achiam2023gpt} models via the API service. 
To achieve more consistent and less random completions, we set the temperature parameter to $0$. 
To enable performance comparisons between humans and \acp{LLM}, we implement both coding approaches in an annotation task, where the models and human coders code all data segments using a human-generated codebook as a common benchmark for evaluation.
We design a prompt for each coding task to instruct the models to perform the coding given an interview segment and the codebook, following the scheme in~\autoref{fig:devices-prompt}. 
We experiment with two different models and prompt engineering techniques, including few-shot learning~\cite{openai_prompt, white2023prompt, brown2020language}:
\begin{itemize}
    \item \textit{GPT-3.5 Turbo vs. GPT-4:} We compare the performance of OpenAI's GPT-3.5 and GPT-4 models. While GPT-4 offers broader general knowledge via a larger context window and can follow complex instructions, GPT-3.5 is offered at a lower price, and versions of this model are also accessible via OpenAI's free AI system ChatGPT.
    \item \textit{Zero-shot vs. One-shot vs. Few-shot:} Providing the model with expected inputs and outputs can increase performance, and teach the model to adopt a specific style~\cite{brown2020language}. In the zero-shot setting, we do not include examples in the prompt. In the one-shot setting, we provide the model with one exemplary data segment and expected output. In the few-shot setting, we provide three examples\footnote{As examples, we use synthetic interview data generated with GPT-4 and annotated by human coders.}.

\end{itemize}

\paragraph{Evaluation of Agreement}
For each coding task, we calculate the \acf{IRR} between the two human coders as well as between the final, unified human annotations and both models via Cohen's Kappa~\citep{cohen1960coefficient, mchugh2012interrater}.
While the \ac{IRR} between human coders is based on the 80\% of annotations that were not used to generate the codebook, we compute the \ac{IRR} between models and humans based on all annotations.
\section{Results: Agreement between Humans and LLMs}
\autoref{tab:no-descr} presents the \ac{IRR} between human coders' and models' annotations.
For tasks~A and~B, human coders achieved almost perfect agreement and for Task~C they reached substantial agreement.
GPT-4 consistently outperforms its predecessor in all three tasks. 
It achieves almost perfect agreement with human annotations on Task~A in all settings -- comparable to inter-human agreement. 
In contrast, with zero or one example, GPT-3.5  only achieves substantial agreement in Task~A, which can be raised to almost perfect agreement when three examples are provided.
In Task~B, both models show moderate to substantial agreement. 
However, in Task~C, the agreement for GPT-3.5 is only fair for the one-shot and few-shot settings, while GPT-4 achieves moderate agreement across all settings.
As the complexity of the coding tasks increases from Task~A to Task~C, there is both a notable decrease in inter-human agreement and a widening gap between inter-human and model-human scores.

\begin{table}[h]
\caption{\ac{IRR} in the form of Cohen's $\kappa$ between the two human coders (top) and between both models and human coders (bottom). Scores for Cohen's $\kappa$ range from $-1$ to $1$, and can be characterized as indicating no agreement ($\kappa < 0.0$), slight agreement ($0.0 \leq \kappa \leq 0.2$), fair agreement ($0.2 < \kappa \leq 0.4$), moderate agreement ($0.4 < \kappa \leq 0.6 $), substantial agreement ($0.6 < \kappa \leq 0.8$), or almost perfect agreement ($0.8 < \kappa \leq 1$)~\cite{landis1977measurement, mchugh2012interrater}.
\label{tab:no-descr}}
\resizebox{\textwidth}{!}{%
\begin{tabular}{l ccc ccc ccc}
\toprule
\multicolumn{1}{c}{} & \multicolumn{3}{c}{\textbf{Task A} (Internet-connected devices)} & \multicolumn{3}{c}{\textbf{Task B} (Apps, programs, services)} & \multicolumn{3}{c}{\textbf{Task C} (Trusted sources)} \\ 
\cmidrule(lr){2-4} \cmidrule(lr){5-7} \cmidrule(lr){8-10}
\multicolumn{1}{l}{inter-human} & \multicolumn{3}{c}{0.97} & \multicolumn{3}{c}{0.83} & \multicolumn{3}{c}{0.63} \\ 
\cmidrule(lr){2-4} \cmidrule(lr){5-7} \cmidrule(lr){8-10}
 & Zero-Shot & One-Shot & Few-Shot & Zero-Shot & One-Shot & Few-Shot & Zero-Shot & One-Shot & Few-Shot \\ 
\cmidrule(lr){2-2} \cmidrule(lr){3-3} \cmidrule(lr){4-4} \cmidrule(lr){5-5} \cmidrule(lr){6-6} \cmidrule(lr){7-7} \cmidrule(lr){8-8} \cmidrule(lr){9-9} \cmidrule(lr){10-10}
gpt-3.5-turbo & 0.74 & 0.78 & 0.85 & 0.60 & 0.58 & 0.56 & 0.43 & 0.36 & 0.34 \\
gpt-4 & 0.97 & 0.94 & 0.95 & 0.75 & 0.74 & 0.75 & 0.56 & 0.55 & 0.58 \\
\bottomrule
\end{tabular}%
}
\end{table}

Providing the model with examples did not significantly improve \ac{IRR}.
However, it played a critical role in mitigating certain challenges, particularly for GPT-3.5, given its inherent limitations, such as restricted input and output sizes.
In particular, few-shot learning helped reduce the number of formatting errors that required manual correction. 
It also played a key role in reducing the number of codes introduced by the model that did not exist in the given codebook, commonly referred to as hallucinations~\cite{ji2023survey}. 
Notably, GPT-3.5 introduced as many as $47$ new, incorrect codes for Task~B -- an amount that could be reduced by more than half with the provision of examples. 
In contrast, GPT-4 generated only two incorrect codes in the same setting.
\section{Discussion}

\paragraph{Inter-human vs. model-human performance}
Although inter-human agreement is in one case equivalent to, and generally higher than model-human agreement, our research highlights the shared challenges faced by human coders and \acp{LLM} when confronted with increasingly complex \ac{QDA} tasks.
Our quantitative findings underscore the inherent difficulty in coding with longer codebooks and contexts, which necessitates capturing nuanced signals and latent themes.
Thus, we advocate for the evaluation of \acp{LLM} on a task-specific basis, recognizing that not all tasks are uniformly compatible with \acp{LLM}.
Various factors, including data collection methods (\eg, survey data, interview data) and segmentation techniques, contribute to the complexity of \ac{QDA}. 
Our ongoing work delves into these dimensions to gain a comprehensive understanding of the challenges and nuances associated with using \acp{LLM} for qualitative coding.

\paragraph{Model choice and few-shot learning}
GPT-4 outperformed GPT 3.5 on all tasks, indicating an improved ability to understand and encode qualitative data. 
While GPT-4 shows higher agreement with human coders, GPT-3.5 offers a considerably lower cost, with input tokens priced 60 times less in our experiments~\cite{openai_cost}.
This makes GPT-3.5 particularly suitable for rapid prototyping, \eg, when engineering prompts for \ac{QDA}.
While few-shot learning did not improve the models' performance, it could play a crucial role in mitigating hallucinations and formatting errors, especially for less capable models such as GPT-3.5.
However, researchers will need to consider whether to use synthetic examples or to manually annotate a subset of data to provide examples. 
This decision may impact performance, calling for a careful analysis of the choice of examples for few-shot learning.
\paragraph{Methodological challenges}
While \ac{IRR} provides a quantitative measure of agreement, it falls short of capturing the depth and subtleties of qualitative analysis~\cite{McDonald2019_inter_rater}. 
In our future work, we aim to conduct a qualitative assessment of our results, along with a detailed error analysis. 
It is critical to recognize that the human benchmark is not infallible, as human coders are susceptible to subjectivity and potential bias~\cite{seaman99, ortloff2023different}. 
This raises the normative question of whether to hold \acp{LLM} to the standard of human coding or to explore alternative scales and benchmarks that might better capture the strengths and limitations of these models.
\paragraph{Risks and limitations}
While \acp{LLM} exhibit remarkable performance in processing and generating language \cite{brown2020language, ziems2023can}, there are inherent limitations in their ability to understand and encode complex contexts, especially in tasks that require a deeper understanding, potentially leading to oversights and misinterpretations.
Moreover, the role and integration of  \acp{LLM} in \ac{QDA} requires a normative discussion. 
For some, the researcher's subjectivity is integral to \ac{QDA}~\cite{mbaleka2020}, where they essentially serve as the measurement instrument~\cite{merriam2015qualitative}. 
In this view, coding involves becoming familiar with the data, and its meaning is derived through human interpretation. 
Using \acp{LLM} for coding could disrupt this process, removing human involvement in text interpretation. 
Thus, aside from determining the suitability of \acp{LLM} for specific coding tasks, it is crucial to reconsider the fundamental purpose of \ac{QDA}.
In addition, the current lack of transparency in \acp{LLM} makes it difficult to track the reasoning behind the results generated.
Small changes in prompt phrasing can result in completely different outputs~\cite{jones2022capturing}. 
Memory issues and limited context windows raise concerns about the consistency and reproducibility of model outputs, as the model does not retain information about its coding decisions between segments.
In addition, the use of \acp{LLM} involves the processing of significant amounts of data, raising questions about the confidentiality and protection of sensitive information of participants, especially when using ChatGPT as a non-API consumer application.
As \acp{LLM} become more integrated into qualitative research, ethical considerations regarding these challenges and their potential impact on research findings become paramount. 
Continued attention to ethical guidelines and responsible use of \acp{LLM} is essential to maintaining the integrity and reliability of qualitative research.

\bibliographystyle{ACM-Reference-Format}
\bibliography{acmart}

\pagebreak
\appendix

\section{PROMPT DESIGN}

\begin{figure}[h]
    \begin{spacing}{1.55}
    \flushleft
    \hltop{You are an expert in qualitative research methods}{Persona}{lightcoral}
    and you are conducting a study on
    \hltop{users' privacy perceptions}{Study Topic}{lightcornflowerblue}.
    You conducted \hltop{semi-structured interviews}{Data Type}{lightfuchsiapink} and want to perform \hltop{thematic analysis}{Methodology}{lightseagreen}. 
    The guiding research question is \hltop{``What technical devices, which are connected to the Internet, do our participants use?''}{Research Question}{lightgreen}.\\
    Identify all \hltop{Internet-connected devices used by the interviewee}{Codes}{lightsalmon} in the given interview transcript and assign them one of the following themes:
    \hltop{[Laptop, Desktop PC, Smartphone, Tablet, Smart Watch, Smart TV, Smart Speaker, Internet Radio, Video Game Console,}{Codebook}{lightgray}\\
    \hl{Home Server, Network-Attached Storage (NAS), Printer, Vacuum Robot, Smart Kitchen Appliances, Smart Toothbrush, Smart Switches and Light Bulbs, Smart Home Devices, Connected Car]}\\
    Format the response as a \hltop{simple list including the themes present in the transcript}{Format}{lightpastelpurple}. If none are found, an empty list [] is returned
    \textit{(Optionally: Examples)}.    
    \end{spacing}
    \caption{Prompt template for coding Internet-connected devices with optional examples. We modify the highlighted text sections to change the coding task.}
    \label{fig:devices-prompt}
\end{figure}

\end{document}